\documentclass[doublecol]{epl2} 
\usepackage{graphicx,xspace,units,subfigure}
\usepackage{float}
\usepackage{amsmath}
\usepackage{amssymb}
\usepackage[normalem]{ulem}
\usepackage{appendix}
\usepackage{epsfig}
\input{epsf}

\newcommand{\noi}{\noindent}

\newcommand{\be}{\begin{equation}}
\newcommand{\ee}{\end{equation}}

\title{Generalized Elastic Model: thermal vs non-thermal initial conditions. Universal scaling, roughening, ageing and ergodicity.}
\shorttitle{Title} %Insert here a short version of the title if it exceeds 70 characters

\author{Alessandro Taloni\inst{1,2}, Aleksei Chechkin\inst{1,3} \and Joseph Klafter\inst{1}}

\institute{                    
  \inst{1} School of Chemistry, Tel Aviv University, Tel Aviv 69978, Israel\\
  \inst{2} CNR-IENI, Via R. Cozzi 53,
      20125 Milano, Italy\\
  \inst{3} Akhiezer Institute for Theoretical Physics, NSC KIPT,
  Kharkov 61108, Ukraine
  }

\abstract{We study correlation properties of the generalized elastic model which accounts for the dynamics of polymers, membranes, surfaces and fluctuating interfaces, among others. We develop a theoretical framework which leads to the emergence of universal scaling laws for systems starting from thermal (equilibrium) or non-thermal (non-equilibrium) initial conditions. Our analysis incorporates and broadens previous results such as  observables' double scaling regimes, (super)roughening and anomalous diffusion, and furnishes a new scaling behavior for correlation functions at small times (long distances). We discuss ageing and ergodic properties of the generalized elastic model in non-equilibrium conditions, providing a comparison with the situation occurring in continuous time random walk. Our analysis also allows to assess which observable is able to distinguish whether the system is in or far from equilibrium conditions in an experimental set-up.}

\begin{document}

\maketitle
\section{Introduction}

Continuum linear systems enjoy an evergreen popularity among the scientific community. This is due to the simplicity of their formulation, to their solvability and to their apparent capability of capturing and reproducing the dynamics of complex natural phenomena. This is the case, for instance, of polymers ~\cite{Doi}, membranes ~\cite{membranes,Zilman}, interfaces ~\cite{Edwards, Krug_1,Krug,Majumdar}, surfaces ~\cite{surfaces}, single-file models ~\cite{Lizana}, all systems that are governed by the following stochastic differential equation 

\begin{equation}
\frac{\partial}{\partial t}\mathbf{h}\left(\vec{x},t\right)=\int d^dx'\Lambda\left(\vec{x}-\vec{x}'\right)\frac{\partial^z }{\partial\left|\vec{x}'\right|^z }\mathbf{h}(\vec{x}',t)+\boldsymbol\eta\left(\vec{x},t\right),
\label{GEM}
\end{equation}

\noi which has been introduced as to generalized elastic model (GEM) ~\cite{Taloni-FLE}. The motion of the $D$-dimensional  stochastic field $\mathbf{h}\left(\vec{x},t\right)$ on the $d$-dimensional substrate $\vec{x}$ is driven by the hydrodynamic interactions, embodied by $\Lambda\left(\vec{r}\right)=1/\left|\vec{r}\right|^\alpha$, by the fractional Laplacian $\frac{\partial^z}{\partial\left|\vec{x}\right|^z}:=-\left(-\nabla^2\right)^{z/2}$~\cite{Samko}, and by the thermal noise random source $\boldsymbol\eta\left(\vec{x},t\right)$ fulfilling the fluctuation-dissipation relation (FD), i.e. $\langle\eta_{j}\left(\vec{x},t\right)\eta_{k}\left(\vec{x}',t'\right) \rangle
=  2k_BT\Lambda\left(\vec{x}-\vec{x}'\right)\delta_{j\,k}\delta(t-t')$. We hereby consider only systems satisfying $z>d$, for which the interface is called rough ~\cite{Krug_1}. Moreover,  systems with local (or screened) hydrodynamic interactions are characterized by $\Lambda\left(\vec{r}\right)\equiv\delta\left(\vec{r}\right)$.

However, the system's dynamics and the ensuing macroscopic observables strongly depends on the initial conditions of the equation (\ref{GEM}). For instance, assuming  the system  in a stationary state at $t=0$  corresponds to take a polymer in a coiled relaxed configuration before starting any measurement. The same can be thought for a floating membrane or a surface, or a rough interface, which has reached the thermal equilibrium condition much before than any experimental observation led off. We will refer to these situations as \emph{thermal initial conditions}. On the other hand, the scientific interest is often directed towards the relaxational properties of the system under study: a common self-question which an experimentalist (or a theorist) asks is ``how do i know/characterize the behavior of a system which is relaxing to its equilibrium configuration?''. Think, for instance, to the polymer relaxational dynamics after the translocation through a nanochannel or a nanoslit, to the evolution of a membrane from a flat initial condition, to the growth of an interface on a flat substrate, to the diffusional properties of an assembly of 1-dimensional Brownian particles equally spaced at $t=0$. These initial conditions go under the name of    
\emph{non-thermal initial conditions} and are characterized by $\mathbf{h}\left(\vec{x},0\right)=0$ $\forall \vec{x}$ ~\cite{Taloni-FLE}. Moreover non-thermal fluctuations, often referred to as ``cytoskeletal fluctuations'', have been shown to be responsible for the non-equilibrium dynamics in many biological systems, such as traveling waves due to protein activity in a flexible membrane, as well as in micropipet experiments using an activated membrane surface (see ~\cite{Chatto} and references therein).  The aim of this letter is to show how the scaling properties of a generic  physical observable depends  on the chosen initial condition, whether thermal or non-thermal. 

\noi Mathematically the question is well-posed. In the case of thermal initial conditions we assume that the system reached the stationarity at  $t=-\infty$: the natural consequence is the use of Fourier transform in time and space, i.e. $\mathbf{h}\left(\vec{q},\omega\right)=\int_{-\infty}^{+\infty}d^dx \int_{-\infty}^{+\infty}dt\,
\mathbf{h}\left(\vec{x},t\right)\,e^{-i\left(\vec{q}\cdot\vec{x}-\omega t\right)}$. The Fourier-Fourier transform of the thermal noise indeed reads $\langle\eta_{j}\left(\vec{q},\omega\right)\eta_{k}\left(\vec{q}',\omega'\right) \rangle=2k_BT(2\pi)^{d+1}\delta_{j\,k}A\left|\vec{q}\right|^{\alpha-d}
\delta\left(\vec{q}+\vec{q}'\right)\delta(\omega+\omega')$, where $A=\frac{(4\pi)^{d/2}}{2^{\alpha}}\frac{\Gamma\left((d-\alpha)/2\right)}{\Gamma\left(\alpha/2\right)}$ if $\frac{d-1}{2}<\alpha<d$. Instead, in case of non-thermal initial conditions we will make use of the Laplace transform in time: $\mathbf{h}\left(\vec{q},s\right)=\int_{-\infty}^{+\infty}d^dx \int_{0}^{+\infty}dt\,
\mathbf{h}\left(\vec{x},t\right)\,e^{-i\left(\vec{q}\cdot\vec{x}\right)-s t}$, then the noise Fourier-Laplace transform is given by $\langle\eta_{j}\left(\vec{q},s\right)\eta_{k}\left(\vec{q}',s'\right) \rangle=2k_BT(2\pi)^{d}\delta_{j\,k}A\left|\vec{q}\right|^{\alpha-d}\frac{\delta\left(\vec{q}+\vec{q}'\right)}{s+s'}$. The local hydrodynamic situation is  achieved by setting formally $A=const$ and $\alpha=d$ in the corresponding long-ranged hydrodynamic
expressions; however, this substitution is not to be intended as a limit ~\cite{Taloni-FLE, Taloni-PRE}.  The results of our analysis lead to the emergence of an universal scaling framework, and ensuing time scale, characterizing the observables' behaviours in both thermal and non-thermal initial conditions. Within this context, we analyze different scaling regimes displayed by the correlation functions for systems presenting  long range or local hydrodynamic interactions. Furthermore,  we discuss  the ageing and ergodic properties of a system during its non-equilibrium relaxational phase.

\section{Two-point two-time $\mathbf{h}$-correlation function}

We hereby derive the following two-point two-time correlation function for the stochastic field component $h_j\left(\vec{x},t\right)$

\be
\begin{array}{l}
\langle\left[h\left(\vec{x},t\right)-h\left(\vec{x},0\right)\right]\left[h\left(\vec{x}',t'\right)-h\left(\vec{x}',0\right)\right]\rangle=\\
\        \ \langle \delta_{t} h\left(\vec{x},t\right)
\delta_{t'} h\left(\vec{x}',t'\right)\rangle,
\label{general_2P_2T_CF}
\end{array}
\ee

\noi where we implicitly dropped the index $j$. We aim at furnishing its scaling expression for systems characterized by long range or local hydrodynamic interactions, whether or not the initial conditions are taken to be stationary. 

\subsection{Thermal initial conditions}

Our starting point is the  solution of the  GEM (\ref{GEM}) in its Fourier representation for thermal initial conditions, which reads $h\left(\vec{q},\omega\right)=\eta\left(\vec{q},\omega\right)/\left(-i\omega+A\left|\vec{q}\right|^{z+\alpha-d}\right)$. Hence, using the Fourier transform of the noise correlation function we derive

\be
\begin{array}{l}
\langle h\left(\vec{q},\omega\right)h\left(\vec{q}',\omega'\right) \rangle_{th}=2k_BT(2\pi)^{d+1}\frac{A\left|\vec{q}\right|^{\alpha-d}}{\omega^2+A^2\left|\vec{q}\right|^{\gamma}}\times\\
\delta\left(\vec{q}+\vec{q}'\right)\delta(\omega+\omega'),
\label{CF_FT_th}
\end{array}
\ee

\noi where $\gamma=2(z+\alpha-d)$. This expression will play a central role in the analysis that 
will follow. After applying the inverse Fourier transform in space and time, we get

\be
\begin{array}{l}
\langle h\left(\vec{x},t\right)
h\left(\vec{x}',t'\right)\rangle_{th}=\frac{k_BT}{(2\pi)^{d/2}}\left|\vec{x}-\vec{x}'\right|^{1-d/2}\times\\
\    \ \int_0^{\infty}d\left|\vec{q}\right|\,\left|\vec{q}\right|^{d/2-z}J_{d/2-1}\left(\left|\vec{q}\right|\left|\vec{x}-\vec{x}'\right|\right)e^{-A\left|\vec{q}\right|^{\gamma/2}\left|t-t'\right|}
\label{CF_th}
\end{array}
\ee

\noi  where $J_{d/2-1}(r)$ represents the Bessel function of order $d/2-1$.  Inserting  (\ref{CF_th}) into the definition (\ref{general_2P_2T_CF}), the two-point two-time correlation function can be arranged in the following  general compact form 

\be
\begin{array}{l}
\langle \delta_{t} h\left(\vec{x},t\right)
\delta_{t'} h\left(\vec{x}',t'\right)\rangle_{th}=\frac{k_BT}{(2\pi)^{d/2}}\left|\vec{x}-\vec{x}'\right|^{z-d}\times\\
\left\{f\left[\frac{t}{\tau\left(\left|\vec{x}-\vec{x}'\right|\right)}\right]+f\left[\frac{t'}{\tau\left(\left|\vec{x}-\vec{x}'\right|\right)}\right]-f\left[\frac{\left|t-t'\right|}{\tau\left(\left|\vec{x}-\vec{x}'\right|\right)}\right]\right\}
\label{2P_2T_CF_th}
\end{array}
\ee

\noi where
$\tau\left(\left|\vec{x}-\vec{x}'\right|\right)=\frac{\left|\vec{x}-\vec{x}'\right|^{\gamma/2}}{A}$ is defined as the correlation time of the distance $\left|\vec{x}-\vec{x}'\right|$ ~\cite{Taloni-PRE}. The scaling function $f\left[u\right]$ is given by

\be
f\left[u\right]=\int_0^{\infty}dy\, y^{d/2-z}J_{d/2-1}\left(y\right)\left(1-e^{-y^{\gamma/2}u}\right)
\label{f_def_th}
\ee

\noi and will be the subject of the upcoming investigation. As a matter of fact, the analysis of (\ref{f_def_th}) reveals two different scaling behaviors whether  $u\ll 1$ or $u\gg 1$. Analyzing the scaling of the correlation function is not a mere mathematical exercise, since in real experiments the scattering data of polymers, membranes or rough surfaces require an expression of eq.(\ref{2P_2T_CF_th}) at  small and large   times or respectively, for long and short distances ~\cite{Majaniemi, Krug, Tong, Sinha}. To show this, we first manipulate  (\ref{f_def_th}) to get 

\be
f\left[u\right]=\int_0^u du'u'^{\frac{d-2(\alpha+1)}{\gamma}}\int_0^{\infty}dy\, y^{\alpha-d/2}J_{d/2-1}\left(\frac{y}{u'^{2/\gamma}}\right)e^{-y^{\gamma/2}}.
\label{f_th}
\ee

\noi For $u\ll 1$, systems characterized by long range hydrodynamic interactions behave unlike systems where hydrodynamic interactions are purely local. In the former case, we can safely replace the exponential in (\ref{f_th}) by 1, achieving

\be
f\left[u\right]\simeq2^{\alpha-d/2}\frac{\Gamma\left(\frac{\alpha}{2}\right)}{\Gamma\left(\frac{d-\alpha}{2}\right)}u
\label{f_hydro_sht_th}.
\ee

\noi On the other hand, when the hydrodynamic interactions are completely screened, the exponential  in (\ref{f_th}) must be expanded to the first order. The ensuing integral can be computed  yielding as a result

\be
f\left[u\right]\simeq2^{z+d/2-2}\frac{z}{\pi}\sin\left(\frac{z\pi}{2}\right)\Gamma\left(\frac{z}{2}\right)\Gamma\left(\frac{z+d}{2}\right)u^2
\label{f_local_sht_th},
\ee

\noi when $z\neq 2m$ ($m\in \mathbb{N}$). If $z = 2m$, the scaling function is exponentially small so that (up to constant factors) 

\be
f\left[u\right]\propto u^{\beta+1}e^{-u^{d-z}},
\label{f_local_marg_sht_th} 
\ee

\noi this result extends to a general system the cases  $z=2,4$ and $d=1$ studied in ~\cite{Majaniemi}. The novelty of eq.(\ref{f_hydro_sht_th}) and (\ref{f_local_sht_th}) and their experimental relevance  will be highlighted in the next section. In the opposite limit $u\gg 1$, there is no difference among systems with long range or local hydrodynamic interactions: this can be seen by expanding to the first order the Bessel function and calculating the ensuing integral, the result reads

\be
f\left[u\right]\simeq\frac{2^{1-d/2}}{z-d}\frac{\Gamma(1-\beta)}{\Gamma\left(\frac{d}{2}\right)}u^\beta
\label{f_lot_th},
\ee

\noi where $\beta=2(z-d)/\gamma$, $0 < \beta < 1$ ~\cite{Taloni-FLE}. We notice that, thanks to the definition of the correlation time $\tau\left(\left|\vec{x}-\vec{x}'\right|\right)$, the long time limit expression (\ref{f_lot_th}) entails a cancellation of the spatial dependence of the two-point two-time correlation function (\ref{2P_2T_CF_th}), i.e. $\langle \delta_{t} h\left(\vec{x},t\right)
\delta_{t'} h\left(\vec{x}',t'\right)\rangle_{th}\simeq\langle \delta_{t} h\left(\vec{x},t\right)
\delta_{t'} h\left(\vec{x},t'\right)\rangle_{th}$, where

\be
\langle \delta_t h\left(\vec{x},t\right)
\delta_{t'} h\left(\vec{x},t'\right)\rangle_{th}=K\left[t^\beta+t'^\beta-\left|t-t'\right|^\beta\right]
\label{2P_2T_CF_lot_th}
\ee

\noi and $
K=\frac{2k_BT\pi^{d/2}}{(2\pi)^d\Gamma\left(d/2\right)}\frac{A^{\beta}\Gamma\left(1-\beta\right)}{z-d}$ as already pointed out in ~\cite{Krug_1,Taloni-FLE}. Therefore the characteristic time $\tau\left(\left|\vec{x}-\vec{x}'\right|\right)$ can be seen as the time up to which the dynamics of two distinct probes in $\vec{x}$ and $\vec{x}'$ is uncorrelated, indeed for $t\gg\tau\left(\left|\vec{x}-\vec{x}'\right|\right)$ the autocorrelation function (\ref{2P_2T_CF_lot_th}) coincides with the two-point two-time correlation function (\ref{2P_2T_CF_th}). Alternatively, one can define the correlation length $\xi(t)=(At)^{2/\gamma}$ ~\cite{Krug,Krug_1,Majumdar,Majaniemi,Tong,Sinha,Lopez,Lopez_1,Lafouresse, Mata, Family} and say that two probes are correlated when $\xi(t)$ exceeds the distance $\left|\vec{x}-\vec{x}'\right|$.

\subsection{Non-thermal initial conditions}

When the GEM (\ref{GEM}) starts from non-thermal initial conditions, its solution in the Fourier-Laplace space is given by $h\left(\vec{q},s\right)=\eta\left(\vec{q},s\right)/\left(s+A\left|\vec{q}\right|^{z+\alpha-d}\right)$ from which it follows

\be
\begin{array}{l}
\langle h\left(\vec{q},s\right)h\left(\vec{q}',s'\right) \rangle_{nth}=
2k_BT(2\pi)^{d}
\frac{\delta\left(\vec{q}+\vec{q}'\right)}{s+s'}\times\\
\          \ \frac{A\left|\vec{q}\right|^{\alpha-d}}{\left(s+A\left|\vec{q}\right|^{\gamma}\right)\left(s'+A\left|\vec{q}\right|^{\gamma}\right)} ,
\label{CF_FT_Nth}
\end{array}
\ee

\noi thanks to expression for the Laplace transform of the noise correlation function. Making an inverse Fourier-Laplace transform in space and time yields

\be
\begin{array}{l}
\langle \delta_{t} h\left(\vec{x},t\right)
\delta_{t'} h\left(\vec{x}',t'\right)\rangle_{nth}=\frac{k_BT}{(2\pi)^{d/2}}\left|\vec{x}-\vec{x}'\right|^{z-d}\times\\
\left\{f\left[\frac{t+t'}{\tau\left(\left|\vec{x}-\vec{x}'\right|\right)}\right]-f\left[\frac{\left|t-t'\right|}{\tau\left(\left|\vec{x}-\vec{x}'\right|\right)}\right]\right\}.
\label{2P_2T_CF_Nth}
\end{array}
\ee

\noi Thus we trace the previous analysis leading to the expressions (\ref{f_hydro_sht_th}), (\ref{f_local_sht_th}), (\ref{f_local_marg_sht_th}) and (\ref{f_lot_th}), in the limit $u\ll 1$ ($t\pm t'\ll \tau\left(\left|\vec{x}-\vec{x}'\right|\right)$) and $u\gg 1$ ($t\pm t'\gg \tau\left(\left|\vec{x}-\vec{x}'\right|\right)$) respectively. In particular $\langle \delta_{t} h\left(\vec{x},t\right)
\delta_{t'} h\left(\vec{x}',t'\right)\rangle_{nth}\sim\langle \delta_{t} h\left(\vec{x},t\right)
\delta_{t'} h\left(\vec{x},t'\right)\rangle_{nth}$ for $t\pm t'\gg \tau\left(\left|\vec{x}-\vec{x}'\right|\right)$, where ~\cite{Krug_1,Taloni-FLE}

\be
\langle \delta_t h\left(\vec{x},t\right)
\delta_{t'} h\left(\vec{x},t'\right)\rangle_{nth}=K\left[(t+t')^\beta-\left|t-t'\right|^\beta\right]
\label{2P_2T_CF_lot_Nth}
\ee

  As an example, let us discuss the situation of local hydrodynamic interactions $\left(\Lambda\left(\left|{\vec r}\right|\right)=\delta\left(\left|{\vec r}\right|\right)\right)$ and  $z=2$, $d=1$ and $A=1$ in (\ref{GEM}). This corresponds  to the equation for Rouse polymers ~\cite{Doi}, for the Edward-Wilkinson chains ~\cite{Edwards}, for the attachment-detachment diffusion model of fluctuating interfaces ~\cite{surfaces}, for single-file systems  ~\cite{Lizana}, and it is also known as 1D diffusion-noise equation ~\cite{VKampen}. In this case we can compute the  expressions (\ref{f_def_th}) exactly: $
f[u]=\sqrt{2u}e^{-\frac{1}{4u}}-\sqrt{\frac{\pi}{2}}\,erfc\left(\sqrt{\frac{1}{4u}}\right)$, 
 where $erfc$ denotes the complementary error function. Expanding $erfc$ for small and large arguments gives $f[u]\simeq (2u)^{3/2}e^{-\frac{1}{4u}}$ if $u\ll1$, and $f[u]\simeq \sqrt{2u}$ if $u\gg 1$, which corresponds to eq.(\ref{f_local_marg_sht_th}) and eq.(\ref{f_lot_th}) respectively ~\cite{Lizana,Majaniemi,Tong}.

\section{Two-point one-time correlation function}

We now turn to the analysis of the following correlation function

\be
\begin{array}{l}
\langle\left[h\left(\vec{x},t\right)-h\left(0,t\right)\right]\left[h\left(\vec{x}',t\right)-h\left(0,t\right)\right]\rangle=\\
\        \ \langle \delta_{x} h\left(\vec{x},t\right)
\delta_{x'} h\left(\vec{x}',t\right)\rangle.
\label{general_2P_1T_CF}
\end{array}
\ee

\noi Our intent is to show the space scaling properties of the GEM (\ref{GEM}) at a given time $t$.

\subsection{Thermal initial conditions}

If the entire system is  in a stationary state,  from (\ref{CF_FT_th}) we can derive  the  structure factor or power spectrum ~\cite{Lopez}:

\be
\langle h\left(\vec{q},t\right)h\left(\vec{q}',t\right) \rangle_{th}=k_BT(2\pi)^{d}\frac{\delta\left(\vec{q}+\vec{q}'\right)}{\left|\vec{q}\right|^{z}}.
\label{spectrum_th}
\ee

\noi  Making an inverse Fourier trasform in space, the previous expression gives

\be
\begin{array}{l}
\langle h\left(\vec{x},t\right)h\left(\vec{x}',t\right) \rangle_{th}=k_BT\frac{\left|\vec{x}-\vec{x}'\right|^{1-d/2}}{(2\pi)^{d/2}}\times\\
\       \ \int_0^{\infty}d\left|\vec{q}\right|\left|\vec{q}\right|^{d/2-z}J_{d/2-1}\left(\left|\vec{q}\right|\left|\vec{x}-\vec{x}'\right|\right),
\label{CF_spectrum_th}
\end{array}
\ee

\noi which is also obtainable from eq.(\ref{CF_th}) by setting $t=t'$. This allows to recast the expression (\ref{general_2P_1T_CF}) as follows

\be
\begin{array}{l}
\langle \delta_{x} h\left(\vec{x},t\right)
\delta_{x'} h\left(\vec{x}',t\right)\rangle_{th}=\\
\frac{k_BT}{(2\pi)^{d/2}}\left\{\ell_{th}\left[\vec{x}\right]+\ell_{th}\left[\vec{x}'\right]-\ell_{th}\left[\vec{x}-\vec{x}'\right]\right\}.
\label{2P_1T_CF_th}
\end{array}
\ee

\noi The analytical form of the function $\ell_{th}\left[\vec{x}\right]$ is

\be
\begin{array}{l}
\ell_{th}\left[\vec{x}\right]=\frac{2^{1-d/2}}{\Gamma(d/2)}\left|\vec{x}\right|^{z-d}\int_0^{\infty}dy\,y^{d-z-1}\times\\
\    \ \left\{1-\Gamma(d/2)\left(\frac{y}{2}\right)^{1-d/2}J_{d/2-1}\left(y\right)\right\}.
\label{l_def_th}
\end{array}
\ee

\noi The analysis of (\ref{l_def_th}) reveals that  the integral in $y$ is convergent if the condition $z<d+2$ is satisfied. As a consequence, systems satisfying the condition $z<d+2$, are characterized by a correlation function (\ref{2P_1T_CF_th}) which is that of a fractional Brownian motion, with the time replaced by the spatial coordinate $\left|\vec{x}\right|$, and the Hurst exponent $H_x = (z-d)/2$, as firstly reported in ~\cite{Majumdar}. On the other hand, if  $z\geq d+2$, the divergence for $y\to 0$ requires the introduction of a lower cut-off $\frac{2\pi\left|\vec{x}\right|}{L}$, where $L$ represents the maximum size of the system. This entails $\ell_{th}\left[\vec{x}\right]\propto L^{z-d-2}\left|\vec{x}\right|^2$ which, plugged into eq.(\ref{2P_1T_CF_th}), points out  that the route mean square difference $\sqrt{\langle \left[h\left(\vec{x},t\right)-h\left(\vec{x}',t\right)\right]^2\rangle_{th}}$ is proportional to the distance $\left|\vec{x}-\vec{x}'\right|$ for \emph{any} system for which $z\geq d+2$, as already known for bending-energy dominated membranes ~\cite{Zilman} and for fluctuating interfaces ~\cite{Majumdar}.

\noi Now, recalling the definition  of the correlation time $\tau\left(\left|\vec{x}\right|\right)$, we can summarize the obtained results  by recasting the two-point one-time correlation function (\ref{2P_1T_CF_th}) in the same fashion as the two-point two-time correlation function (\ref{2P_2T_CF_th}). This can be done by expressing $ \ell_{th}\left[\vec{x}\right]=\left|\vec{x}\right|^{z-d}g_{th}\left[\frac{\tau(L)}{\tau\left(\left|\vec{x}\right|\right)}\right]$, with $ g_{th}[u]=g_{th}=const$ if $z<d+2$, and $g_{th}[u]\propto u^{2(z-d-2)/\gamma}$ if  $z>d+2$.
Thus, the morphological transition occurring at $z=d+2$ ~\cite{Krug,Majumdar},  turns out to be a general property satisfied also  by systems with long range hydrodynamics such as membranes, polymers or viscoelastic surfaces. In analogy with fluctuating interfaces we will refer to the systems for which  $z<d+2$ as Family-Vicsek systems, and to the systems fulfilling $z\geq d+2$ as super-rough systems ~\cite{Lopez, Lopez_1}.

\subsection{Non-thermal initial conditions}

In case of non-thermal initial conditions, the correlation function $\langle h\left(\vec{x},t\right)h\left(\vec{x}',t\right) \rangle_{nth}$ can be obtained from  the power spectrum

\be
\langle h\left(\vec{q},t\right)h\left(\vec{q}',t\right) \rangle_{nth}=k_BT(2\pi)^{d}\frac{\delta\left(\vec{q}+\vec{q}'\right)}{\left|\vec{q}\right|^{z}}\left(1-e^{-2A\left|\vec{q}\right|^{\frac{\gamma}{2}}t}\right)
\label{spectrum_Nth}
\ee

\noi  which follows from eq.(\ref{CF_FT_Nth}). Making an inverse Fourier transform in space we get 

\be
\langle h\left(\vec{x},t\right)h\left(\vec{x}',t\right) \rangle_{nth}=\frac{k_BT}{(2\pi)^{d/2}}\left|\vec{x}-\vec{x}'\right|^{z-d}
f\left[\frac{2t}{\tau\left(\left|\vec{x}-\vec{x}'\right|\right)}\right]
\label{CF_spectrum_Nth}.
\ee

\noi due to the definition (\ref{f_def_th}). Alternatively, the expression (\ref{CF_spectrum_Nth}) follows immediately from  eq.(\ref{2P_2T_CF_Nth}) setting $t=t'$. Anyhow, the quantity (\ref{CF_spectrum_Nth}) is of experimental importance for the interpretation of scattering data, indeed it is straightforwardly connected to the dynamic structure factor. By instance, in case of rough surfaces ~\cite{Majaniemi,Tong}, a widely used form of (\ref{CF_spectrum_Nth}) for short times (long distances) is exponential, in agreement with the expression (\ref{f_local_marg_sht_th}) ~\cite{Sinha}. However, we point out that this behaviour represents only a marginal situation ($z=2m$) for systems characterized by local hydrodynamic interactions.  As a matter of fact, the expression (\ref{f_local_sht_th}) yields a more general  result: the long distance behavior of  (\ref{CF_spectrum_Nth}), for any $z\neq 2m$, exhibites a decay $\propto\left|\vec{x}-\vec{x}'\right|^{-z-d} $. On the other hand, eq.(\ref{f_hydro_sht_th}) furnishes a testable new prediction for the scattering data coming from long ranged hydrodynamics viscoelastic systems.  

\noi In analogy with the thermal case, the two-point one-time correlation function (\ref{general_2P_1T_CF}) can be casted as

\be
\begin{array}{l}
\langle \delta_{x} h\left(\vec{x},t\right)
\delta_{x'} h\left(\vec{x}',t\right)\rangle_{nth}=\frac{k_BT}{(2\pi)^{d/2}}\left\{\ell_{nth}\left[\vec{x},t\right]+\right.\\
\left.\ell_{nth}\left[\vec{x}',t\right]-\ell_{nth}\left[\vec{x}-\vec{x}',t\right]\right\}
\label{2P_1T_CF_Nth}
\end{array}
\ee

\noi where 

\be
\begin{array}{l}
\ell_{nth}\left[\vec{x},t\right]=\frac{2^{1-d/2}}{\Gamma(d/2)}\left|\vec{x}\right|^{z-d}\int_0^{\infty}dy\,y^{d-z-1}\times\\
\    \ \left\{1-\Gamma(d/2)\left(\frac{y}{2}\right)^{1-d/2}J_{d/2-1}\left(y\right)\right\}\left(1-e^{-y^{\gamma/2}\frac{2t}{\tau\left(\left|\vec{x}\right|\right)}}\right).
\label{l_def_nth}
\end{array}
\ee

\noi  After the following change of  variable $v=y\left(\frac{2t}{\tau\left(\left|\vec{x}\right|\right)}\right)^{2/\gamma}$, the short and long time analysis of eqs.(\ref{l_def_nth}) and (\ref{2P_1T_CF_Nth}) can be performed. For short times, $t\ll\tau\left(\left|\vec{x}\right|\right)$, to the leading order we get

\be
\begin{array}{l}
\langle \delta_{x} h\left(\vec{x},t\right)
\delta_{x'} h\left(\vec{x}',t\right)\rangle_{nth}\sim K(2t)^\beta,
\label{2P_1T_CF_sht_Nth}
\end{array}
\ee

\noi implying the cancellation of the spatial dependence in (\ref{2P_1T_CF_Nth}). Eq.(\ref{2P_1T_CF_sht_Nth}) is exactly the mean square displacement of a tracer (probe particle) in a generic point $\vec{x}$, when system starts from non-thermal initial conditions ~\cite{Taloni-FLE}.  
A different situation occurs whenever one is concerned with the long time  limit, $t\gg\tau\left(\left|\vec{x}\right|\right)$, of (\ref{l_def_nth}). First, we express $\ell_{nth}\left[\vec{x},t\right]=\left|\vec{x}\right|^{z-d}g_{nth}\left[\frac{t}{\tau\left(\left|\vec{x}\right|\right)}\right]$, then we cast the function $g_{nth}\left[u\right]$ as

\be
g_{nth}\left[u\right]\simeq
\begin{array}{ccc}
g_{th}-const\times u^{2\frac{z-d-2}{\gamma}}   &   &  z<d+2\\
u^{2\frac{z-d-2}{\gamma}}   &   &  z\geq d+2
\label{g_nth}
\end{array}
\ee

\noi This quantity is intimately connected to the emergence of the so-called \emph{anomalous scaling} or \emph{anomalous roughening} ~\cite{Krug, Lopez, Lopez_1}. This arises whenever the \emph{local} width $w^2\left(l,t\right)=\langle\langle\left[ h\left(\vec{x},t\right)-\langle h\left(\vec{x},t\right)\rangle_{l^d}\right]^2\rangle_{l^d}\rangle$ scales differently from the \emph{global} width $W^2\left(L,t\right)=\langle\langle\left[ h\left(\vec{x},t\right)-\langle h\left(\vec{x},t\right)\rangle_{L^d}\right]^2\rangle_{L^d}\rangle$. For $t\geq\tau(L)$ the global width reaches the value  $W^2\left(L,t\right)\propto L^{2\chi}$, where $\chi=\frac{z-d}{2}$ is the roughness exponent ~\cite{Family}, which coincides with the spatial Hurst exponent $H_x$ ~\cite{Mata}. On the other hand, the local width scaling law is given by $w^2\left(l,t\right)\sim l^{2\chi_{loc}}$ for $\tau\left(l\right)\ll t\ll\tau\left(L\right)$. The local roughness exponent $\chi_{loc}$ can be, in general, different from $\chi$:  whenever it happens, a system is said to present anomalous roughening. The anomalous scaling has been detected in a large collection of experiments and models (see ~\cite{Lopez,Lopez_1,Lafouresse,Mata} and reference therein). In our framework it stems from the scaling properties of the function $g_{nth}$, indeed  $w^2\left(l,t\right)\simeq l^{z-d}g_{nth}\left[\frac{t}{\tau\left(l\right)}\right]$, which gives $\chi_{loc}=\chi$ for Family-Vicsek systems and $\chi_{loc}=1$  for super-rough systems  ~\cite{Lopez}.  Our analysis extends this result to long range hydrodynamics systems, showing that anomalous roughening is a general phenomenon not only restricted to the domain of rough surfaces. By instance, flexible Zimm polymers  ~\cite{Doi} ($z=2,\alpha=1/2,d=1$ in eq.(\ref{GEM})) are believed to fall in the Family-Vicsek universality class, while fluid membranes and  semiflexible polymers \cite{membranes,Zilman} ($z=4,\alpha=1$ and $d=2$ and $1$, respectively)  should exhibit anomalous superrough scaling.  As a matter of fact, the correlation function (\ref{general_2P_1T_CF}) has been studied in ~\cite{Zilman} for membranes starting from thermal initial conditions, but its behavior has never been addressed in case of non-thermal initial conditions.

\section{Time vs ensemble average}

Our analysis now turns to the effects that the specific initial conditions may have on the time  average of an observable, which is function of a probe particle trajectory only, i.e. of a tracer particle placed at position $\vec{x}$. In general, it is of a very broad  interest to assess the correctness of the ergodic hypothesis in real systems, i.e. if the time and ensemble average coincide. In the specific, it might be  extremely useful for an experimentalist to establish whether the time average of a one-point observable can be representative of the thermodynamic state of the entire system, if in or out of equilibrium.
Firstly, we define the  observable to be the  square displacement 
$\delta_t^2h\left(\vec{x},t\right)=\left[h\left(\vec{x},t\right)-h\left(\vec{x},0\right)\right]^2$. We then define the time average over a trajectory of length $\mathcal{T}$, as

\be
\overline{\delta_t^2h\left(\vec{x},t\right)}^\mathcal{T}=
\frac{1}{\mathcal{T}-t}\int_0^{\mathcal{T}-t}dt_0\,\left[h\left(\vec{x},t+t_0\right)-h\left(\vec{x},t_0\right)\right]^2.
\label{MSD_time_def}
\ee

\noi In general we expect that (\ref{MSD_time_def}) is a fluctuating quantity dependent on the chosen stochastic trajectory: it is then legitimate to take its ensemble average  $\langle\overline{\delta_t^2h\left(\vec{x},t\right)}^\mathcal{T}\rangle$ and compare it with $\langle\delta_t^2h\left(\vec{x},t\right)\rangle$ ~\cite{weakEB_Yossi}.

\subsection{Thermal initial conditions}

 If the system is in thermal equilibrium, the probe dynamics is  ergodic, indeed the stochastic motion of a probe particle in $\vec{x}$ is governed by a fractional Langevin equation with stationary fractional Gaussian noise ~\cite{Taloni-FLE}, then $\lim_{\mathcal{T}\to\infty}\overline{\delta_t^2h\left(\vec{x},t\right)}^\mathcal{T}=\langle\delta_t^2h\left(\vec{x},t\right)\rangle_{th}=2Kt^\beta$ (Fig.\ref{fig.1}(a)).

\subsection{Non-thermal initial conditions}

In order to calculate the ensemble average of (\ref{MSD_time_def}), i.e. $\langle\overline{\delta_th\left(\vec{x},t\right)}^\mathcal{T}\rangle_{nth}$,   we need to get the probe MSD between times $t+t_0$ and $t_0$: $\langle
\left[h\left(\vec{x},t+t_0\right)-h\left(\vec{x},t_0\right)\right]^2\rangle_{nth}$. From (\ref{2P_2T_CF_lot_Nth}) it is immediate to obtain

\be
\begin{array}{l}
\langle\left[h\left(\vec{x},t+t_0\right)-h\left(\vec{x},t_0\right)\right]^2\rangle_{nth}=\\
K\left\{[2(t+t_0)]^\beta+(2t_0)^\beta+2t^\beta-2(t+2t_0)^{\beta}\right\}.
\label{MSD_shift_Nth}
\end{array}
\ee

\noi 
At first, we notice that the MSD exhibits ageing since it displays a strong dependence on $t$ and $t_0$ when both are large ~\cite{Eli}. We then  study its behavior for $t_0\ll t$ and $t_0\gg t$

\be
\langle\left[h\left(\vec{x},t+t_0\right)-h\left(\vec{x},t_0\right)\right]^2\rangle_{nth}\simeq
\begin{array}{ccc}
K(2t)^\beta &  & t_0\ll t,\\
2Kt^\beta &  & t_0\gg t.
\label{MSD_CF_nth}
\end{array}
\ee

\begin{figure}
\begin{center}
\epsfig{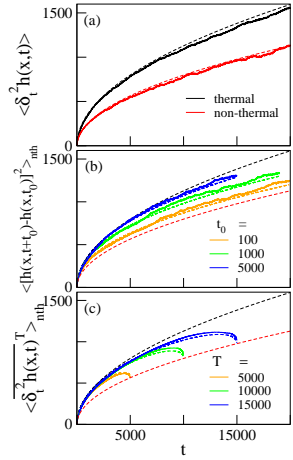}
\end{center}
\caption{(Color online) MSD of a tagged tracer in a single-file system ($z=2$, $d=1$ and $A=1/\zeta$, with $\zeta=1.0$ the damping ~\cite{Lizana}). Simulations are performed integrating the stochastic dynamics of  a file of $N=2000$ Brownian ($k_BT=1.0$) hard-core pointlike particles,  moving  along a ring of length $L=20000$. Thermal initial conditions correspond to pick a random file's configuration from an uniform distribution at $t=0$, whereas non-thermal initial conditions correspond to take particle equally spaced of a distance $L/N$ ($L/N=10$ in the simulation).  Panel (a): MSD ensemble average for thermal (solid upper black line) and non-thermal initial conditions (solid bottom red line), plotted against the theoretical predictions $2Kt^\beta$ and $K(2t)^\beta$ respectively (dashed lines). Panel (b): $\langle\left[h\left(\vec{x},t+t_0\right)-h\left(\vec{x},t_0\right)\right]^2\rangle_{nth}$ for different $t_0$ (solid lines), upper black dashed line corresponds to $2Kt^\beta$, bottom red dashed line to $K(2t)^\beta$, while middle color dashed lines stand for eq.(\ref{MSD_shift_Nth}). Panel (c):  $\langle\overline{\delta^2_t h\left(\vec{x},t\right)}^\mathcal{T}\rangle_{nth}$ for different trajectories's length $\mathcal{T}$. Solid lines show the outcome of the numerical simulations, dashed color lines represent the corresponding theoretical eq.(\ref{MSD_time_T_Nth}). Upper and bottom dashed lines are the same as in panel (b). Simulation results are averaged over $40000$ different probe's trajectories.}
\label{fig.1}
\end{figure}

\noi Therefore, for $t\ll t_0\ll\tau(L)$, where $\tau(L)$ represents the characteristic relaxational time of a system of size $L$, the probe's diffusion is like that for thermal initial conditions, albeit the entire system is in a non-stationary regime. In this condition an experimentalist could not decide whether the system is in equilibrium or not. These results are supported by numerical simulations of a single file system (see Fig.\ref{fig.1}(b)). 
An illustrative description of the obtained results may come from the  comparison of (\ref{MSD_shift_Nth}) with the corresponding quantity of a subdiffusive continuous time random walk (CTRW) process $\xi(t)$. It is known that CTRW MSD exhibits ageing ~\cite{Sokolov,Eli,weakEB_Yossi}. In the same limits of eq.(\ref{MSD_CF_nth}) one has $\langle\left[ \xi(t+t_0)-\xi(t_0)\right]^2\rangle\simeq t^\mu$ for $t_0\ll t$ and $\langle\left[ \xi(t+t_0)-\xi(t_0)\right]^2\rangle\simeq t_0^{\mu-1}t$ for $t_0\gg t$, where $\mu$ is the exponent of the waiting time distribution $\psi(t)\sim t^{-\mu-1}$ ($0<\mu<1$). Thus, the MSD grows subdiffusively in the limit  $t_0/t\to 0$ but normally when $t_0/t\to\infty$, which is at odd with the corresponding limits in (\ref{MSD_CF_nth}), where  the difference is only in the prefactor. Moreover the ageing properties of the MSD in  CTRW are very different from those of (\ref{MSD_shift_Nth}): in particular, in  the regime $t_0\gg t$, diffusion is slowed down in CTRW as $t_0$ increases ~\cite{Eli}, while the probe diffusion in case of non-thermal initial conditions loses the dependence on $t_0$, to the main order.

\noi  We now turn to the ensemble average of (\ref{MSD_time_def}).  Integrating eq.(\ref{MSD_shift_Nth}) one gets

\be
\begin{array}{l}
\langle\overline{\delta_t^2h\left(\vec{x},t\right)}^\mathcal{T}\rangle_{nth}=K\times\\
\left[2t^\beta+\frac{2^\beta\left(\mathcal{T}^{\beta+1}+(\mathcal{T}-t)^{\beta+1}-t^{\beta+1}\right)-(2\mathcal{T}-t)^{\beta+1}+t^{\beta+1}}{(1+\beta)(\mathcal{T}-t)}\right],
\label{MSD_time_T_Nth}
\end{array}
\ee

\noi from which it is apparent the dependence on the trajectory length $\mathcal{T}$. 
\noi This is indeed shown by the outcome of the numerical simulations of single file system, perfectly reproduced by the theoretical prediction (\ref{MSD_time_T_Nth}) (Fig\ref{fig.1}(c)). The natural requirement in experiments is to take the limit $t\ll \mathcal{T}$. We then expand expression (\ref{MSD_time_T_Nth}) for small values of the parameter $t/\mathcal{T}$ obtaining $\langle\overline{\delta_t^2h\left(\vec{x},t\right)}^\mathcal{T}\rangle_{nth}\simeq 2Kt^\beta\left(1-\frac{2^{\beta}-1}{2(1+\beta)}\frac{t}{\mathcal{T}}\right)$. Therefore, if the system is prepared in non-thermal initial conditions, in the limit of $t/\mathcal{T}\to 0$, the ensemble average of   (\ref{MSD_time_def})  tends to the value of the ensemble (or time) averaged MSD at equilibrium, i.e. $\langle\overline{\delta_t^2h\left(\vec{x},t\right)}^\mathcal{T}\rangle_{nth}\to \langle\delta_t^2h\left(\vec{x},t\right)\rangle_{th}$.
Thus, it would be impossible for an experimentalist to assess whether the system is in equilibrum or far from it, just by looking at the probe time averaged MSD when the condition $t\ll\mathcal{T}$ is fulfilled. However, when $t$ approaches $\mathcal{T}$,  the  $\mathcal{T}$-dependence appearing in  (\ref{MSD_time_T_Nth}) would become apparent and it would constitute the signature of the non-equilibrium thermodynamical state of the entire system (Fig\ref{fig.1}(c)). Finally, the expression (\ref{MSD_time_T_Nth}) is in contrast to that obtained in CTRW model, for which  $\langle\overline{\delta_t^2\xi(t)}^\mathcal{T}\rangle\sim \mathcal{T}^{\mu-1}t$ when $t\ll\mathcal{T}$ ~\cite{weakEB_Yossi}.

\section{Conclusions}
In this letter we established a universal theoretical framework for the correlation functions in a generalized elastic model. Within this framework, the scale invariance and the scaling regimes attained by any physical observable emerge naturally, getting a  physical meaning in terms of the correlation time $\tau$. On one hand, we found new scaling behaviors of the correlation functions at small times (large distances) providing testable predictions for the behavior of the dynamic structure factor in case of rough surfaces and viscoelastic systems.   On the other, we showed that anomalous roughening is a physical phenomenon that can be detected also in processes characterized by long-ranged hydrodynamic interactions. Moreover, we analyzed ageing and ergodic properties of the generalized elastic model performing time and ensemble average of the squared displacement. Our results point out that the probe's MSD  attains its equilibrium value although the entire systems is far from it.

\acknowledgments
ACh acknowledges financial support from European Commission via MC
IIF grant No.219966 LeFrac. AT and Ach are grateful to S. Majumdar and E. Barkai for illuminating discussions.

\end{document}